\documentclass[11pt]{article}
\usepackage{times}
\usepackage{cospar}
\usepackage[sectionbib]{natbib}
\usepackage{graphicx}
\usepackage[figuresright]{rotating}

\title{NON-THERMAL X-RAY AND RADIO EMISSION FROM THE SNR N 157B}

\author{John R. Dickel\address{Astronomy Department, University of Illinois 
at Urbana-Champaign, 1002 West Green Street, Urbana IL 61801, USA},
    and
    Shiya Wang$^{1}$}

\begin{document}

\maketitle

\begin{abstract}
The supernova remnant N 157B contains a pulsar and three distinct synchrotron 
components with rather unusual properties.  1)  A somewhat irregular elliptical pulsar wind 
nebula (PWN) visible in both X-ray and radio wavelengths.  The nebula is quite 
symmetrical with an extent of about 10 $\times$ 5 parsecs but offset along 
the long axis by about 4 pc from the pulsar position.  It is apparently the 
result of a short-lived injection of energetic particles, perhaps starting 
at the time of 
explosion.  2)  A very bright X-ray shock region located just outside the pulsar 
position in the edge of the PWN.  This is undetected in the radio.  We
attribute this to a new burst of particles from the pulsar suggesting there
are multiple episodes rather than continuous injection.  3)  The 
beginning of a radio synchrotron shell on the southern side of the SNR where 
thermal X-rays appear to arise suggesting that N 157B is starting to become 
a composite SNR. 
\end{abstract}

\section*{INTRODUCTION}

N 157B (Henize, 1956) is a Crab-type supernova remnant (SNR) just 7 
arcmin from the center of 30 Doradus (Bode, 1801) in the Large Magellanic 
Cloud (LMC).  It contains 
a 16-ms X-ray pulsar undetected at any other wavelength (Marshall 
et al. 1998).  There is bright non-thermal X-ray emission with structure on 
arcsec scales just around the pulsar with an extended feature off to the 
northwest (Wang and Gotthelf, 1998a, 1998b; Wang et al. 2001).  There is 
bright non-thermal radio emission from the extended feature but not at the 
pulsar location (Lazendic et al. 2000).   We shall call the extended emission 
region the pulsar wind nebula (PWN). 
The overall struture suggests that the pulsar is moving 
toward the southeast.  There is also extended radio emission toward the south 
that gives a hint of a shell, suggesting that the remnant may be in transition 
to becoming a composite remnant with a shell and a pulsar wind nebula.

The differences in the radio and X-ray structures plus the apparent 
large motion of the pulsar make this SNR unusual.  We shall describe its 
properties and then discuss the implications of the data.

\section*{X-RAY -- RADIO COMPARISONS}

\subsection{Morphology}

Figures 1 and 2 show the simularities of the radio and X-ray emission of the 
PWN component of N 157B but a striking difference toward 
the pulsar.  The radio emission 
in the pulsar wind component sits on a plateau of the rest of the remnant.  
Fine structure in the PWN appears very similar in both 
wavelength ranges although the radio emission extends further northwest.  
This structure probably represents clumpiness in the interstellar medium or 
in pre-explosion mass loss of the progenitor. 

The peak in the X-ray emission in the compact source 
around the pulsar is 13 times the brightness of the peak in the PWN but in 
the radio there is nothing seen above the residual brightness of the PWN.  
The pulsar lies at $05^{h}37^{m}47.2^{s}$ and 
$-70^{\circ}10'23''$ (Wang and Gotthelf, 1988b) about 16$''$ out from the center 
along the SE axis of the tail.  The pulsar is about 1$''$ closer to the 
center of the tail than the peak of the X-ray emission.

\begin{figure}
\center
\includegraphics[width=9cm,angle=270]{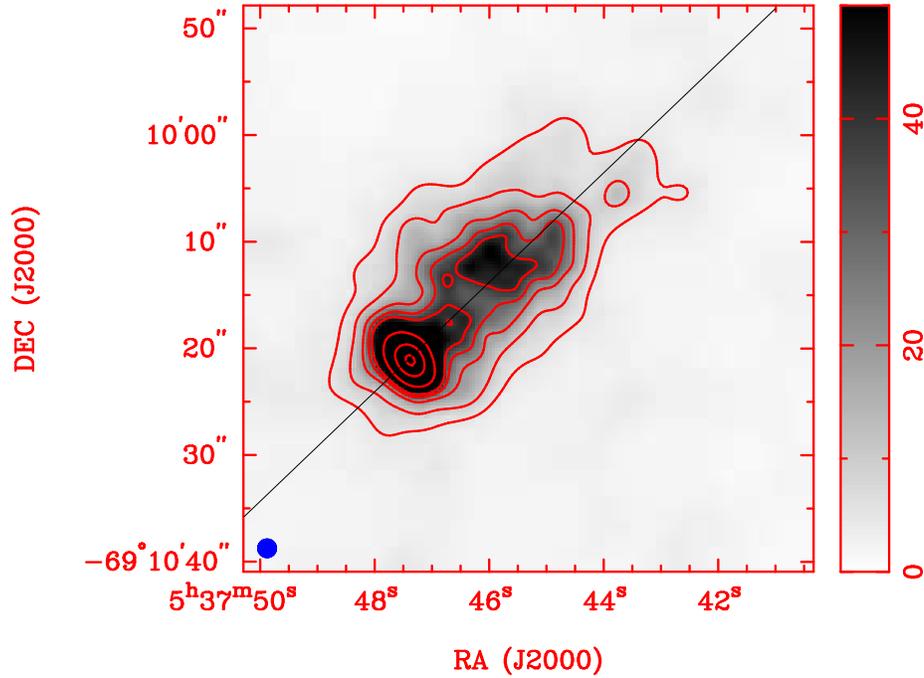}
\caption{Chandra X-ray image of the pulsar wind part of the supernova remnant 
N 157B smoothed to a resolution of 1.8 arcsec.  
The contours are 5, 10, 20, 30, 40, 100, 300, and 600 
counts over the 0.1 -- 10 kev range of the HRC detector.  The thin line shows 
the location of slices shown in Figure 3.}
\end{figure}

\begin{figure}
\center
\includegraphics[width=9cm,angle=270]{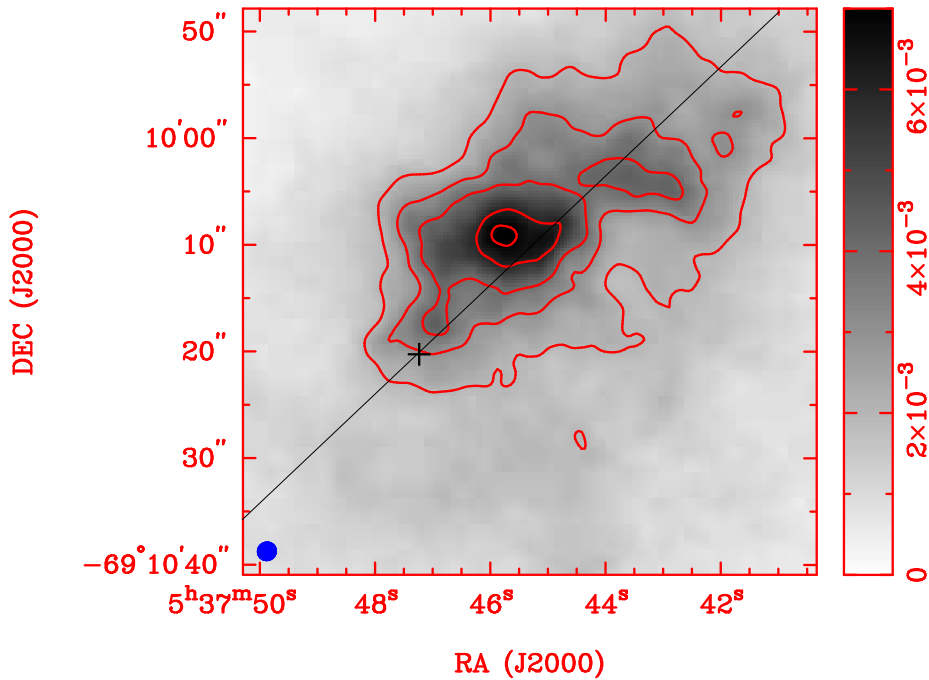}
\caption{ATCA 6-cm radio image of the pulsar wind part of the supernova 
remnant N 157B at a resolution of 1.8 arcsec.  The contours are 2, 3, 4, 6, 
and 7 mJy beam$^{-1}$.  The thin line 
shows the location of slices shown in Figure 3 and the cross is the 
position of the pulsar.}
\end{figure}

\begin{figure}[ht]
\center
\includegraphics[width=90mm,angle=270]{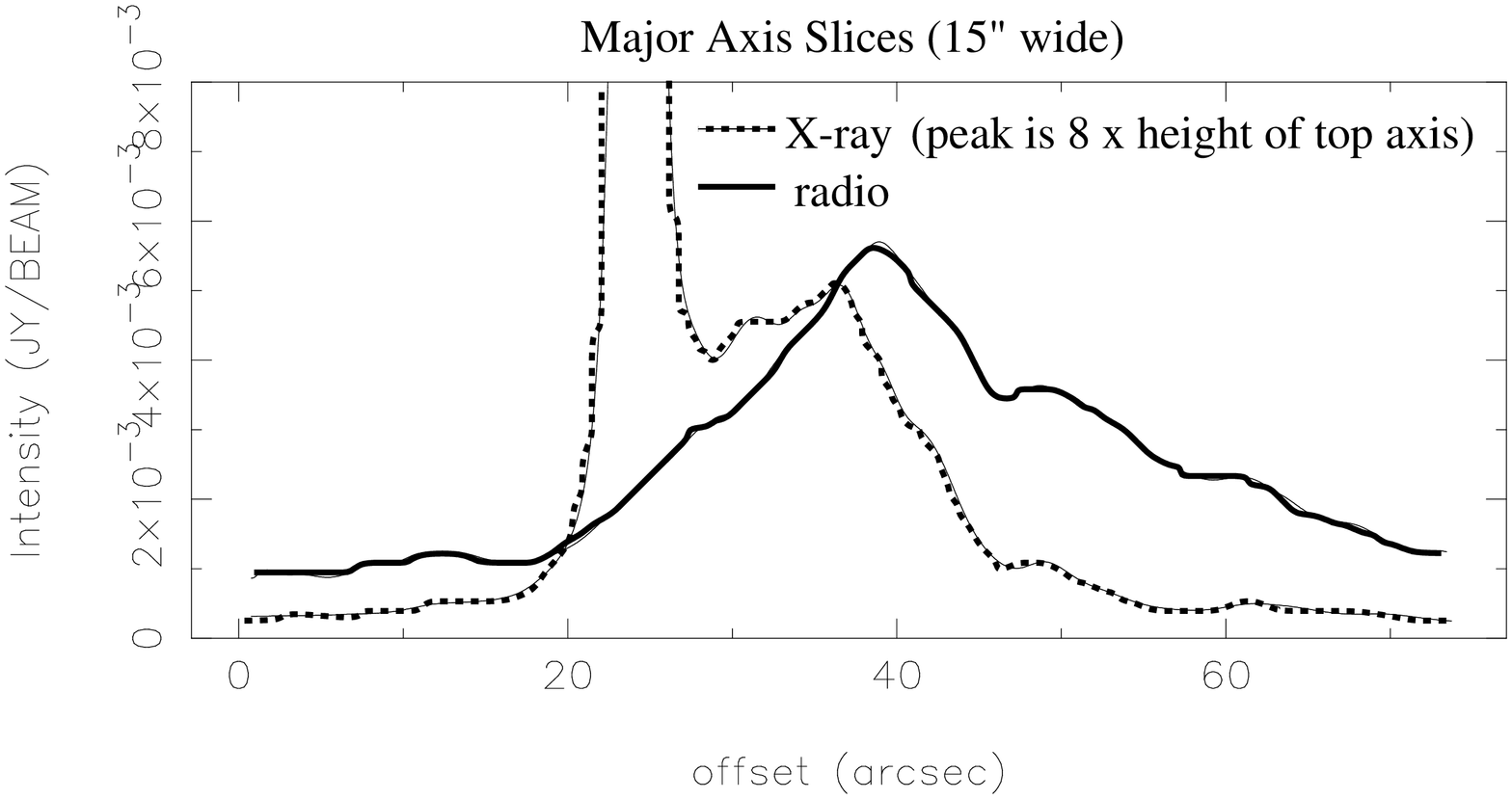}
\caption{X-ray and radio slices from southeast to northwest through the 
pulsar position and along the major 
axis of the elliptical PWN 
of N 157B as shown in Figures 1 and 2. The slices are averaged over a width 
of 15$''$.  The radio units are Jy beam$^{-1}$ and the X-ray ones are counts 
psf$^{-1}$ normalized to fit on the same scale.}
\vspace*{-0.1cm}
\center
\includegraphics[width=90mm,angle=270]{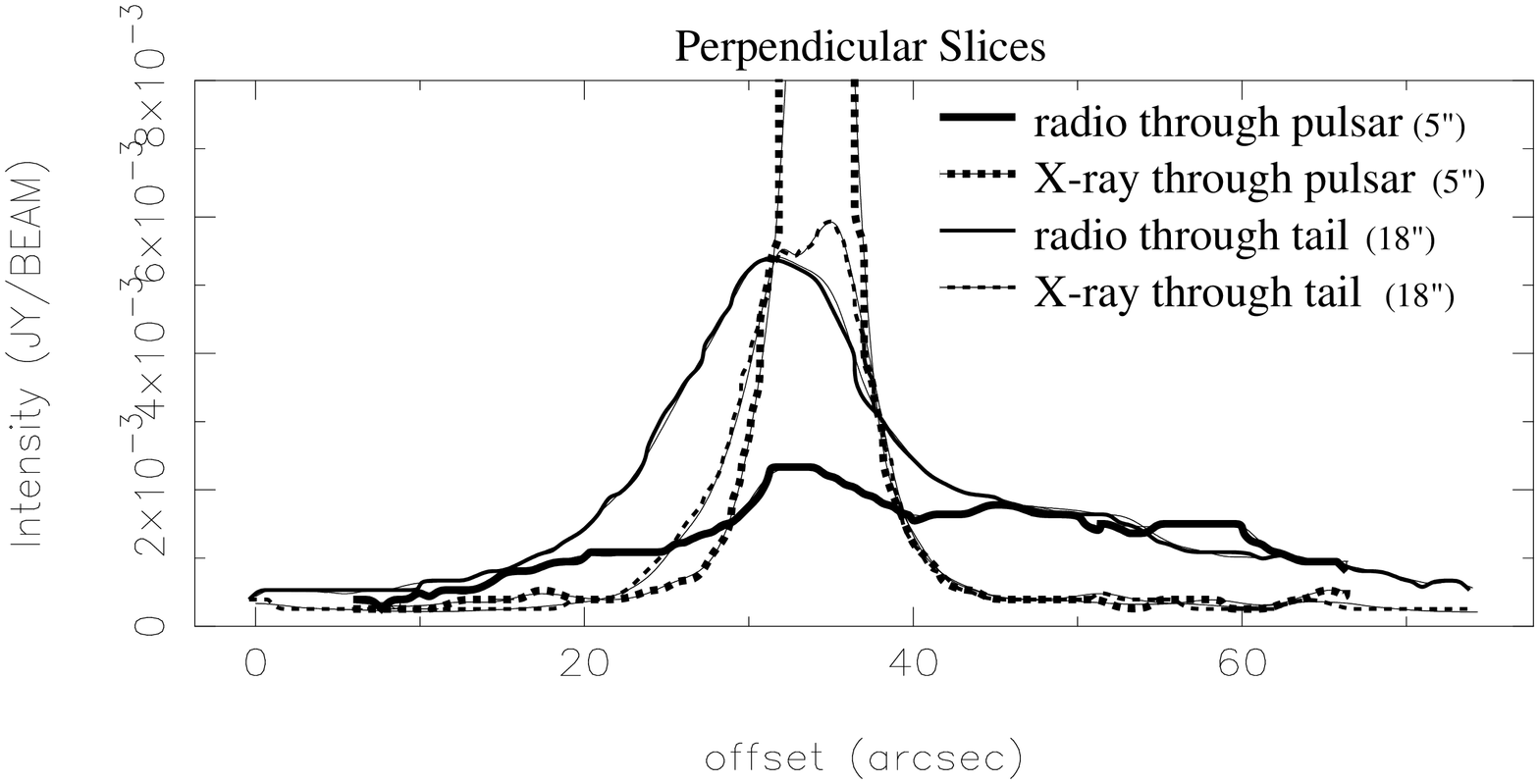}
\caption{Slices from northeast to southwest perpendicular to the major axis 
shown in Figures 1 and 2.  The ones through the center of the 
PWN are 18$''$ wide centered 
on the radio peak and the ones through the pulsar position are 5$''$ wide.}
\end{figure}

\clearpage

Further details of the emission are revealed by the slices presented in 
Figures 3 and 4. The emission from the PWN is clearly more extended in all 
directions at radio wavelengths than at X-ray ones.  The overall symmetry is 
the same, however, with 
an approximately elliptical shape centered about a point at 
$05^{h}37^{m}45.5^{s}$ and $-69^{\circ}10'09''$.  The radio emission falls off 
uniformly out to a semi-major axis extent of 20$''$ in the SE-NW direction and 
10$''$  in the NE-SW direction.

There is no sign of the pulsar or any 
enhancement in the radio emission, $< 0.1$ mJy beam$^{-1}$, at the pulsar's 
position of $05^{h}37^{m}47.2^{s}$ and $-70^{\circ}10'23''$ (Wang and 
Gotthelf, 1988b). 
The non-thermal X-rays around the pulsar position, 
on the other hand, show a strong approximately elliptical component, 
about 7$''$ $ \times $ 3$''$ with its long axis perpendicular 
to the long axis of the PWN tail.  Wang and Gotthelf (1998) suggested that this
small source could be a bow-shock from the particles leaving the moving pulsar.
We shall 
henceforth call that structure the shock region. From the inner (NW) 
edge of this shock, the X-ray emission first decreases somewhat  and then 
increases gradually toward the 
radio center of the PWN but peaks 3$''$ before the radio and then falls 
sharply 
toward the northwest.

\subsection*{Spectra}

To compare the actual brightnesses of the features, we show their spectra 
in Figure 5.  The squares represent the integrated values for the radio 
emission of the entire SNR.  They give a spectral index, $\alpha$, of  
$-0.19 \pm 0.1$, where the flux density $S_{\nu}$ $\propto$  $\nu^{\alpha}$ 
(Lazendic et al. 2000).  XMM-Newton spectra, that cannot resolve angular 
detail, show that most of the X-ray emission from the SNR has a steep 
power-law spectrum with $\alpha = -1.83 \pm 0.03$ although some thermal 
emission is present as well (Dennerl et al. 2001).  They do not give a value 
for the actual X-ray flux.

For the PWN, the lower frequency 
radio data do not have sufficient resolution for a good separation of the 
components so we report only the 4.8- and 8.6-GHz results (Lazendic et al. 
2001).  The spectral index for the PWN is more uncertain because of the 
SNR background.  The value of $\alpha = -0.1 \pm 0.2$ could easily be the 
same as that of the whole SNR.  The error of the spectral index for this fit to 
only the two data points tries to take into account the uncertainty in 
evaluation of the background.  We cannot determine the radio spectrum of the 
shock region because it is not detected.  We do show the upper limits for its 
flux density at the two radio frequencies.

The x-ray spectra are from the paper by Wang and Gotthelf (1998b).  Their 
formal fits give values of $\alpha = -1.5 ~(+ 0.2, -0.3)$ for the PWN and 
$\alpha = 
-1.9 \pm 0.2$ for the bow-shock.  Realizing that these errors are only for the 
formal fits to the data, we suggest that the slopes of both components could 
be the same but that of the PWN cannot be much steeper than that of the 
shock region. 

\begin{figure}
\center
\includegraphics[width=90mm,angle=0]{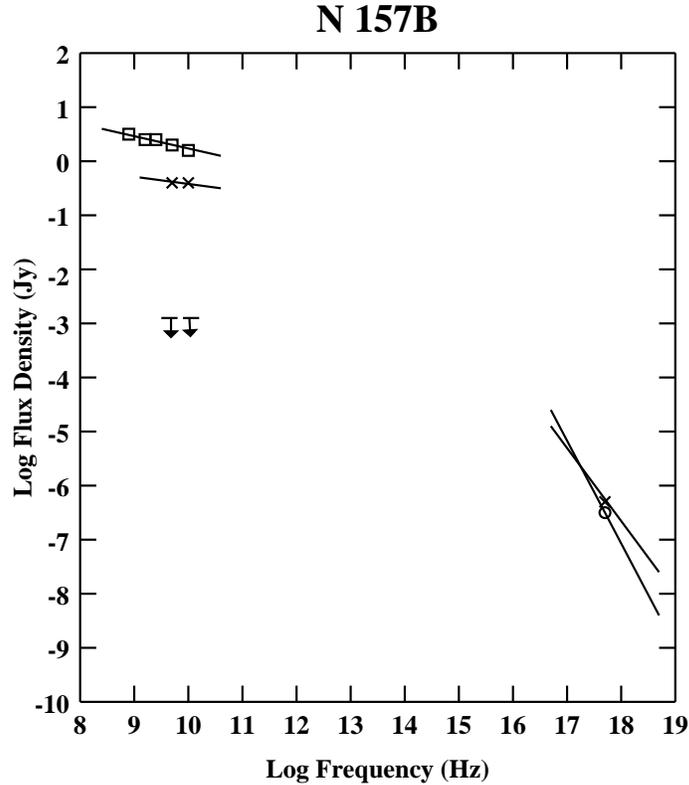}
\caption{Full spectrum of the SNR N157B.  The open boxes are for the entire 
SNR; the crosses are for the pulsar wind nebula; and the open circle and 
upper-limit arrows are for the shock region.}
\end{figure}

\section*{DISCUSSION}

For analysis, we divide 
the SNR into three parts as outlined by Wang and Gotthelf (1998a) and 
discussed above: the  elliptical pulsar wind nebula extending northwest 
from the pulsar with major and minor axes of $10 \times  5$ pc;  the brignt 
shock region ($\sim 2 \times 1/2$ pc) centered just outside the pulsar;  and 
the 
entire SNR, about $30 \times 18$ pc across, which extends well beyond the 
images in Figures 1 and 2.  

Assuming that the pulsar has been moving 
southeastward from an explosion site at the radio peak, we can 
estimate its speed using the characteristic spin-down age of 5000 years 
(Marshall et al. 1998; Wang and Gotthelf, 1998b).  To have moved 4 pc, the 
pulsar has 
a speed of 800 km sec$^{-1}$, large but not impossible for a pulsar 
(Arzoumanian et al. 2002)).  Because the PWN emission falls off in all 
directions from its 
center, including toward the pulsar, we are led to the conclusion that there 
was a brief period of particle injection by the pulsar, lasting up to 
about 1000 years after the explosion occurred.  That period would account for
the $\sim$ 3/4 pc shift of the X-ray peak from the radio one and also allow 
for the apparent aging of the electrons on the outer edge of the PWN relative 
to the center.  As higher energy electrons decay more rapidly from their 
synchrotron emission, those further from the pulsar should be the oldest 
and thus have reduced X-ray 
emission relative to the radio and steeper spectra (Pacini and Salvati, 1973;
Reynolds and Chevalier, 1984).  The particle injection and/or stimulation 
cannot have lasted much longer, however, or the emission in the PWN would 
still be bright toward the southeast between the PWN peak and the current
location of the pulsar. 

The emission from the shock region just around the pulsar appears to have 
arisen from a second injection of particles.  The 
X-ray brightness rises so drastically just there and appears to sit on top of
the decreasing PWN emission in that direction.  The radio emission is below 
detection level.  We also note that if the injection of particles has been
continuous,  the 
X-ray spectrum of the region closest to the pulsar should be flatter (harder)
than that of the region further from the current pulsar position, but that is
not the case.  Thus the newer particles around the pulsar must have been 
injected with a different spectrum or be interacting with a very different 
medium that those injected earlier near the center of the overall PWN.  
Finally, the center of this shock source is 
slightly (about 1/4 pc) outside the pulsar position.  These conditions lead us
to believe that the object is indeed a shock region caused by the
supersonic motion of the particles coming from the pulsar moving at about 800
km sec$^{-1}$ plus their injection speed and perhaps interacting with a 
reverse shock from the intitial blast wave.  Without a more accurate 
separation of the
thermal component of the X-ray emission in that region to evaluate the 
temperature and density of the gas, we cannot determine the actual sound speed
but the relative motion could be enough to generate a strong shock wave and 
compressed
magnetic fields across which to accelerate particles to even greater energies
than they get from the pulsar.  The pulsar may have also encountered a density 
enhancement which will further increase the 
total emission to produce the high observed X-ray brightness.  Indications of a
clumpy medium around N 157B include many H$\alpha$ filaments and a very low
polarization of the radio emission, presumably caused by significant Faraday 
rotation (Lazendic et al. 2000).  The other lumps in the PWN may represent 
temporary increases due to previous brief episodes of acceleration near 
density enhancements.

We don't know the radio spectral index of the shock region so we don't know if
the energy injection in that spectral range is the same for both episodes.  
To get an approximate
ratio of radio to X-ray luminosities, we will assume that the radio spectrum
of the shock region 
is the same as the rest of the SNR, $\sim -0.2$.  Extrapolation of any 
value steeper than about $-$0.3 would require less X-ray emission than 
observed.  With the adopted spectral index, we calculate an upper limit to the 
radio luminosity between 10$^{7}$ and 10$^{11}$ Hz of $\sim 4 \times 10^{32}$ 
erg sec$^{-1}$ or about 1/2000 of the X-ray luminosity in the 0.2 -- 4 keV 
band (Wang et al. 2001).  This upper limit is a low ratio for either a pulsar wind alone,
e.g. 0540-693 has a ratio of radio to X-ray luminosity of 1/120,  or for an SNR
with shock generated X-ray emission, e.g. AD1006 has a ratio of about 1/100 (taken
from Allen et al. 2001).  Perhaps the combined processes may create a harder
spectrum at low energies but then a sharper break at the transition between
the emission in the two frequency ranges.

While still qualitative, this description of multiple episodes of particle
injection from the pulsar is the only one we can come up with to explain
all the data.  Better radio resolution and sensitivity plus improved spectra in
all wavelength bands, including a detection of infrared and optical continuum
emission, will be important for checking these ideas.    

\section*{ACKNOWLEDGEMENTS}

We thank Paul Plucinsky, Steve Reynolds, Brian Fields, You-Hua Chu, Douglas 
Bock, Martin Guerrero, and Pat Slane for help and ideas.  This research is 
supported by NASA grant NAG 5-11159.

\vspace*{0.5cm}

\noindent 
johnd@astro.uiuc.edu, swang9@astro.uiuc.edu  \\

\end{document}